\documentstyle[12pt,a4,epsf]{article}

\makeatletter
\@addtoreset{equation}{section}
\makeatother

\newcommand{\VEV}[1]{\left\langle #1 \right\rangle}

\newcommand{\Mp}{M_P}

\begin{document}
\begin{titlepage}

\begin{flushright}
hep-ph/0107313\\
KUNS-1729\\
\today
\end{flushright}

\vspace{4ex}

\begin{center}
{\large \bf
A natural solution for the $\mu$ problem with anomalous $U(1)_A$ 
gauge symmetry
}

\vspace{6ex}

\renewcommand{\thefootnote}{\alph{footnote}}

Nobuhiro Maekawa\footnote
{e-mail: maekawa@gauge.scphys.kyoto-u.ac.jp
}

\vspace{4ex}
{\it Department of Physics, Kyoto University,\\
     Kyoto 606-8502, Japan}\\
\end{center}

\renewcommand{\thefootnote}{\arabic{footnote}}
\setcounter{footnote}{0}
\vspace{6ex}

\begin{abstract}

Recently we proposed an attractive scenario of grand unified 
theories with anomalous $U(1)_A$ gauge symmetry, in which 
doublet-triplet splitting is naturally realized in $SO(10)$ 
unification using Dimopoulos-Wilczek mechanism and realistic 
quark and lepton mass matrices can be obtained in a simple way.
In this paper we show that there is a mechanism in which 
the doublet Higgs obtains the supersymmetric mass which is 
proportional to the SUSY breaking parameters. This mechanism 
can be applied easily in the above scenario. The point is that 
the mass term, which is forbidden by SUSY zero mechanism, can 
be induced by SUSY breaking. The proportional coefficient is 
controlled by the anomalous $U(1)_A$ charges.

\end{abstract}

\end{titlepage}


Recently we proposed an attractive scenario of supersymmetric
(SUSY)
grand unified theories (GUTs) with anomalous $U(1)_A$ gauge symmetry,
in which doublet-triplet splitting is 
naturally realized in $SO(10)$ unification using 
Dimopoulos-Wilczek mechanism, and realistic quark and lepton
mass matrices can be obtained in a simple way
\cite{maekawa}.

In the scenario, the mass term of the Higgs field is forbidden
by the holomorphy ( SUSY zero mechanism). This is because the anomalous 
$U(1)_A$ charge of the Higgs field is taken to be negative.
However, in order to give a mass to higgsino, the SUSY Higgs mass term
is required. The SUSY Higgs mass $\mu$ must be of oder of the weak
scale, namely, the SUSY breaking scale. This is a mystery in the minimal
SUSY standard model, because at a glance we have no reason that the 
SUSY parameter becomes the same order of the SUSY breaking parameters.
This is called the $\mu$ problem
\cite{mu}.
In the super gravity scenario, there are several natural solutions 
for the $\mu$ problem
\cite{giudice,casas} by using non-renormalization operator 
in the K\"ahler potential (Giudice and Masiero
\cite{giudice}) 
or in the superpotential (Casas and Mu\~nos
\cite{casas}).
However, if the Higgs mass term in tree level is forbidden by 
some symmetry as in our scenario, these mechanisms for the $\mu$ 
problem do not
work well, 
though R-symmetry can be an exceptional one. Since our model has
no R-symmetry, it is important to examine other mechanisms to induce
the supersymmetric Higgs mass term related with the SUSY breaking
scale. There are several other attempts
\cite{singlet,problem,kolda,hempfling,nir}
 to solve the $\mu$ problem. 
One of them is to introduce a light singlet which couples with
the Higgs doublet
\cite{singlet,kolda}. The vacuum expectation
value (VEV) of the singlet field can become of order of the SUSY 
breaking scale, so suitable
$\mu$ parameter is induced. In addition to the problem that the induced
$\mu$ parameter is unstable under radiative correction of heavy particle
and non-renormalizable terms
\cite{problem}, however, in our 
scenario, it is not easy to introduce a light singlet $S$ with positive 
charge (the positive charge is required for the singlet to couple 
to Higgs field), because the mass term of the singlet field with positive
charge is not forbidden.

In this paper, we examine a mechanism which solves the 
$\mu$ problem. The mechanism can be naturally applied to our scenario.
The generated Higgs mass $\mu$ is proportional to the 
SUSY breaking parameters and the coefficient is controlled by anomalous
$U(1)_A$ charges. The point is that since the Higgs mass term is 
forbidden
by the SUSY zero mechanism, when SUSY is broken, the $\mu$ term must
be induced. Note that if all the SUSY breaking parameter become zero, 
the $\mu$ term must vanish. Therefore the $\mu$ parameter must
be proportional to the SUSY breaking parameters. Since all the coefficients
are controlled by anomalous $U(1)_A$ charges, the proportional coefficient
is also determined by the anomalous $U(1)_A$ charges. 

Let us recall the SUSY zero mechanism. First of all, we assume
that $D$-flatness condition of the anomalous $U(1)_A$ gauge symmetry
leads to $\VEV{\Phi}=\lambda\Mp$,
because $D_A=\frac{g_A}{2}(\lambda^2\Mp^2-|\Phi|^2)$, where
$\Phi$ has negative
anomalous $U(1)_A$ charge $\phi=-1$ and $\lambda<1$ (Actually we usually
adopt $\lambda\sim 0.2$ for reproducing the Cabbibo angle.). 
Here $\Mp$ is some gravity scale and usually taken as the reduced Planck 
mass, $1/\sqrt{8\pi G_N}$. In the following, we use the units in which
$\Mp=1$. Then the hierarchical structure
of Yukawa couplings can be obtained as
\begin{equation}
W_Y=\Phi^{q+u+h}QUH\rightarrow \lambda^{q+u+h}QUH,
\end{equation}
if $q+u+h\leq 0$.
Here $q,u$ and $h$ are anomalous $U(1)_A$ charges of 
the superfields $Q,U$ and $H$
\footnote{
Throughout this paper we denote all the superfields and the scalar 
component fields with uppercase letters
and their anomalous $U(1)_A$ charges with the corresponding lowercase 
letters.
}.
The terms with negative total anomalous $U(1)_A$ charge are forbidden 
by
the anomalous $U(1)_A$ gauge symmetry, while the terms with non-negative 
total charge are allowed, because the negative charge of the singlet 
$\Phi$ can compensate for the positive charge, as addressed above.
The vanishing of the coefficients resulting from the anomalous $U(1)_A$ 
gauge 
symmetry is called  SUSY zero mechanism. 
In the previous paper
\cite{maekawa}, this SUSY zero mechanism plays
an essential role to solve the two biggest problems in grand unified theory,
the doublet-triplet splitting problem and hierarchy problem of quark and 
lepton mass matrices.
For example, the Higgs mass term $\mu$ in tree level can be forbidden
by the mechanism if the anomalous $U(1)_A$ charge of the Higgs $H$
is negative.
It is obvious that the vacuum expectation values of the gauge singlet 
operators with
positive anomalous $U(1)_A$ charges must vanish so that the SUSY
zero mechanism works well. On the other hand, the gauge invariant
operator with negative charge can have the VEV. The value is 
written
\begin{equation}
\VEV{O}\sim \lambda^{-o},
\end{equation}
if the $F$-flatness condition determines the VEV
\cite{maekawa}.

Generally if SUSY is broken, the coefficients, which vanish by SUSY zero 
mechanism, become tiny non-zero values which are proportional to the
SUSY breaking parameters. In the followings, we estimate the coefficients
by anomalous $U(1)_A$ charges.

Before examining this mechanism, we try to apply the Giudice-Masiero 
mechanism
to induce the SUSY Higgs mass $\mu$. When SUSY is broken by the $F$-term
$F_T$ of a field $T$, a K\"ahler term 
\begin{equation}
\int d\theta^4 \lambda^{|2h-t|}T^\dagger HH
\label{Giudice}
\end{equation}
induces the SUSY Higgs mass term
\begin{equation}
\int d\theta^2 \mu HH=\int d\theta^2 \lambda^{|2h-t|}F_T^\dagger HH.
\end{equation}
Here $H$ is the Higgs superfield, whose representation is, for example,
 ${\bf 10}$
in the context of $SO(10)$ unification.
We have to take $t=0$ so that the gaugino mass is obtained by
$\VEV{F_T}\sim m_{SB} \Mp$, because the supersymmetric field strength has 
vanishing
anomalous $U(1)_A$ charge. Here $m_{SB}$ is a typical SUSY breaking scale,
which is of order of the weak scale.
Then the induced Higgs mass term has suppression
factor $\lambda^{|2h|}$, so it is much
smaller than the weak scale unless $|2h|\leq 1$
\cite{nir}.

We now examine the solution for the $\mu$ problem in a simple example.
The essential point of this mechanism is that the VEV shift of a heavy
singlet field by SUSY breaking. In the literature
\cite{hall}, it is shown that the SUSY breaking terms produce the VEV shift
of heavy particles of order the SUSY breaking scale
in the context of super gravity scenario.
If there is a heavy singlet which has vanishing VEV in SUSY limit and
couples to the 
Higgs field, then shifting the VEV solves the $\mu$ problem. The argument
is essentially the same as in Ref.\cite{hempfling}, but they requires
R-symmetry, which is not a symmetry in our scenario \cite{maekawa}.
Below we show that such a situation is easily obtained in our scenario,
namely, R-symmetry is not an essential ingredient of the mechanism.

Before examining the detail, we figure out the essence of the mechanism.
We introduce the superpotential $W=\lambda^sS+\lambda^{s+z}SZ$, where 
$S$ and $Z$ are singlet fields with positive anomalous $U(1)_A$ charge $s$ 
and with negative charge $z$, respectively ($s+z\geq 0$). 
Note that the single term of $Z$
is not allowed by SUSY zero mechanism, while usual symmetry cannot forbid
this term. This is an essential point of this mechanism. The SUSY vacuum is 
at
$\VEV{S}=0$ and $\VEV{Z}=\lambda^{-z}$. After SUSY is broken, these 
VEVs are modified. To determine the VEV shift of $S$, which we would 
like to know because the singlet $S$ with positive charge can couple to 
the Higgs field with negative charge, the most important SUSY breaking 
term is the tadpole term of $S$, namely $\lambda^s \Mp^2A S$. Here $A$ 
is a SUSY breaking parameter of order of the weak scale. By this tadpole 
term, the VEV of $S$ appears as $\VEV{S}=\lambda^{-s-2z}A$. If we have 
$\lambda^{s+2h} SH^2$, the SUSY Higgs mass is obtained as 
$\mu=\lambda^{2h-2z}m_{SB}$, which is proportional to the SUSY breaking 
parameter $m_{SB}$ and the proportional coefficient can be
of order 1 if $h\sim z$. 

Let us examine the detail below.
The superpotential is written
\begin{equation}
W=\lambda^{s+2h} SH^2+\lambda^s S+\lambda^{s+z}SZ+\lambda^{2s}S^2,
\end{equation}
where for simplicity, we introduce only the last term for breaking 
the R-symmetry. Introducing the other R-symmetry breaking terms does not
change the following results drastically.
The $F$-flatness conditions are
\begin{eqnarray}
F_S&=&\frac{\partial W}{\partial S}=\lambda^{s+2h} H^2+\lambda^s 
  +\lambda^{s+z}Z+2\lambda^{2s}S=0, 
 \label{F_S} \\
F_Z&=&\frac{\partial W}{\partial Z}=\lambda^{s+z}S=0, 
  \label{F_Z} \\
F_H&=&\frac{\partial W}{\partial H}=\lambda^{s+2h}SH=0.
  \label{F_H}
\end{eqnarray}
Since $S=0$ satisfy the two $F$-flatness conditions $F_Z=F_H=0$,
the VEVs of the other fields $Z$ and $H$ are not fixed completely
by these $F$-flatness conditions. Though the desired VEV is 
$(\VEV{Z},\VEV{H^2})\sim (\lambda^{-z},0)$, the VEV of $H^2$ can
have non-vanishing value. 
(Here only for simplicity, we assume that the $D$-flatness condition 
of the anomalous $U(1)_A$ gauge symmetry determine the VEV 
$\VEV{\Phi}\sim \lambda$. However, in principle, the Froggatt-Nielsen 
field $\Phi$ is a dynamical variable, so we have to resolve the 
$D$-flatness condition in addition to the above $F$-flatness 
conditions. We will discuss this point lator.)

After SUSY is broken, the SUSY breaking terms are given as
\begin{eqnarray}
V_{SB}&=&m_S^2|S|^2+m_Z^2|Z|^2+m_H^2|H|^2 \\
      &+&(\lambda^{s+2h} A_{SH^2}SH^2+\lambda^s A_S S+\lambda^{s+z}A_{SZ}SZ
      +\lambda^{2s}A_{S^2}S^2+h.c.).
      \label{SB}
\end{eqnarray}
Here $m_X$ and $A_Y$ ($X=S,Z,H$ and $Y=SH^2,S,SZ,S^2$) are SUSY breaking
parameters.
If we neglect the $D$-term contribution to the potential,
the potential is obtained by
\begin{equation}
V=\left|\frac{\partial W}{\partial S}\right|^2
  +\left|\frac{\partial W}{\partial H}\right|^2
  +\left|\frac{\partial W}{\partial Z}\right|^2+V_{SB}.
\end{equation}
Here we assume that the K\"ahler potential is minimal one 
$K=|H|^2+|Z|^2+|S|^2$ for simplicity, but more general K\"ahler
potential does not change the following result drastically unless 
the K\"ahler potential has a singularity.
The stationary conditions are written
\begin{eqnarray}
\frac{\partial V}{\partial S}&=&2F_S^\dagger \lambda^s 
 +\lambda^{2s+2z}S^\dagger
 +4\lambda^{2s+4h}|H|^2S^\dagger+m_S^2S^\dagger+\lambda^{s+2h}A_{SH^2}H^2 \\
 &&+\lambda^{s+z}A_{SZ}Z+2\lambda^sA_S=0,    
 \label{1st} \\
\frac{\partial V}{\partial Z}&=&F_S^\dagger \lambda^{s+z}+m_Z^2Z^\dagger
+\lambda^{s+z}A_{SZ}S=0, 
\label{2nd}\\
\frac{\partial V}{\partial H}&=&2F_S^\dagger \lambda^{s+2h}H
+2\lambda^{2s+4h}|S|^2H^\dagger+m_H^2H^\dagger+2\lambda^{s+2h}A_{SH^2}SH=0.
\label{3rd}
\end{eqnarray}
The third condition (\ref{3rd}) leads to the following two cases;
a) $\VEV{H}=0$ and b) $\VEV{H}\neq 0$.
The second condition (\ref{2nd}) determines the $F$ term of the $S$ field
$F_S=-\lambda^{-s-z}m_Z^2Z^\dagger-A_{SZ}S$, which is of order $m_{SB}^2$ 
if
$\VEV{S}\sim O(m_{SB})$.
This is important to induce the correct size of $B$ parameter.
It is easily checked that with vanishing SUSY breaking parameters, 
the above
three conditions become three $F$-flatness conditions $F_S=F_H=F_Z=0$.
In the following, we assume that the vacuum can be expanded as
$\VEV{X}=X_0+X_1+\cdots (X=S,Z,H)$ using the SUSY breaking scale as the 
expansion parameter. Here $X_0$ represent SUSY
vacua, namely, $S_0=0$ and $F_S(S_0,Z_0,H_0)=0$. 
The first condition (\ref{1st}) gives 
\begin{equation}
S_1=-\frac{\lambda^sA_S+\lambda^{s+z}A_{SZ}Z_0}
         {\lambda^{2s+2z}+4\lambda^{2s+4h}|H_0|^2}.
         \label{S1}
\end{equation}

In the case a) ($\VEV{H}=0$), the relation $F_S(S_0,Z_0,H_0)=0$ 
leads to 
$Z_0=\lambda^{-z}$. Then from eq.(\ref{S1}), $S_1$ is given by
$S_1=\lambda^{-s-2z}A$. At the vacuum, the value of
the potential is roughly estimated as 
\begin{equation}
V_a\sim \lambda^{-2z}|A|^2.
\end{equation}

Let us consider the case b) ($\VEV{H}\neq 0$). 
If the VEV $\VEV{Z}>>\lambda^{-z}$, 
the value of the potential $V_b\sim m_Z^2|Z|^2>>V_a$.
Therefore we
take $Z_0\sim O(\lambda^{-z})$. Then the equation $F_S(S_0,Z_0,H_0)=0$ 
leads to $H_0\sim O(\lambda^{-h})$. From the eq. (\ref{S1}), 
\begin{equation}
S_1=-\frac{\lambda^sA_S+\lambda^{s}A_{SZ}}
         {\lambda^{2s+2z}+4\lambda^{2s+2h}}
         \sim \left(\begin{array}{c} \lambda^{-s-2z}A \quad (z\leq h) \\
                                     \lambda^{-s-2h}A \quad (z\geq h).
                    \end{array}\right.
\end{equation}
Then 
$2F_S^\dagger \lambda^{s+2h}+4\lambda^{2s+4h}|S_1|^2+m_H^2
+2\lambda^{s+2h}A_{SH^2}S_1=0$ and $F_S(S_0,Z_0,H_0)=0$ determine
the $H_0$ and $Z_0$ definitely. Since all VEVs are determined, the value 
of the potential at the minimum can be estimated as
\begin{equation}
V_b\sim \left(\begin{array}{c} \lambda^{-2z} |A|^2 (z\leq h) \\
                               \lambda^{-2h}|A|^2 (z\geq h).
              \end{array}\right.
\end{equation}
Therefore $z>h$ leads to $V_a>V_b$, namely the desired vacuum (case a) 
becomes local minimum. On the other hand, if $z\leq h$, we obtain
$V_a\sim V_b$. Therefore, the desired vacuum (case a) can be global 
minimum though it is dependent on the $O(1)$ coefficients.

Below we focus on the desired vacuum (case a) even if the vacuum is local
minimum. Then the VEV of $S$ induces the supersymmetric Higgs mass term
$\mu $ as
\begin{equation}
\lambda^{s+2h}SH^2\rightarrow \lambda^{s+2h}\VEV{S}H^2=\lambda^{2h-2z}A H^2.
\end{equation}
It is interesting that the $\mu$ term is proportional to the SUSY breaking
parameter $A$. The proportional coefficient is determined by the anomalous
$U(1)_A$ charges as $\lambda^{2h-2z}$.
When $h\sim z$, we can obtain the natural scale of the SUSY Higgs mass
$\mu$. 
 The Higgs mixing term $B\mu$ can be obtained from the SUSY term
$\lambda^{s+2h}SH^2$ and the
SUSY breaking term $\lambda^{s+2h}A_{SH^2}SH^2$ as
$\lambda^{s+2h}F_S\sim \lambda^{2h-2z}m_{SB}^2$ and 
$\lambda^{2h-2z}A^2\sim \mu A$, respectively.
Therefore the relation $B\sim m_{SB}$ is naturally obtained
\footnote{
If doublet-triplet splitting is realized by fine-tuning or some
accidental cancellation, the Higgs mixing
$B\mu$ can become intermediated scale $m_{SB} M_X$ as discussed in
Ref.\cite{kawamura}, where $M_X$ is the GUT scale.
However, once the 
doublet-triplet splitting is naturally solved  as in 
Ref.\cite{maekawa}, such a problem disappears }.
This is a solution for the $\mu$ problem. Note that the condition 
$h\sim z$ can be satisfied because both fields $H$ and $Z$ have
 negative charges. 

At a glance, requiring the condition $h\sim z$ is artificial. However,
recall that even the Giudice-Masiero mechanism requires an additional
condition $h\sim 0$.

In the above argument, we almost fix the VEV $\VEV{\Phi}\sim \lambda$, 
which is considered to be determined by the $D$-flatness condition of the 
anomalous $U(1)_A$ gauge symmetry. Since the Froggatt-Nielsen field 
$\Phi$ is a dynamical variable, in principle, we have to reconsider
the $D$-flatness condition of anomalous $U(1)_A$ gauge symmetry to 
determine the VEVs. 
However, the result is almost the same as that discussed in my paper
if $\VEV{\Phi}\sim \lambda$ and $z<-1$. Since the VEV of $Z$ is 
$\lambda^{-z}$, which is much smaller than the VEV of $\Phi$, 
reconsidering D-flatness condition $\lambda^2-|\Phi|^2-z|Z|^2=0$ makes 
only a tiny shift in the VEV of $\Phi$.
Of course there is another possiblity that the other vacuum appears,
for example, $\VEV{\Phi}=0$ and $\VEV{Z}\sim \lambda$ (case c). 
(Actually this vacuum
satisfies all the $F$-flatness conditions (\ref{F_S}), (\ref{F_Z}) and
(\ref{F_H}).)  In such a case, the role of
$\Phi$ is exchanged for that of $Z$. The condition $\phi=-1>h$ leads to 
$V_c\sim {\rm min}(\lambda^{2h/z}|A|^2, \lambda^2m_{SB}^2)>
\lambda^{-2h}|A|^2\sim V_b$, 
which means that the vacuum c is an only local minimum.
Moreover, if we adopt the charges $s=-n z$ 
($n$ is a positive integer), then the $F$-flatness condition of $S$ 
requires  $\Phi^{s}+Z^n\sim 0$, namely, $\Phi\sim Z^{-1/z}>\lambda$, 
which is inconsistent with $D$-flatness condition. Then the vacuum c is
not allowed. In any cases, the situation is not changed drastically by
examining the $D$-flatness condition.  This is consisitent
with  the number of equations and variables. If we add one
equation($D$-flatness condition) and one variable ($\Phi$), it is 
expected that the number of the vacua does not increase drastically.

In this paper, we have focused on the $\mu$ problem. However, the mechanism
can be applied to more general case. We can apply this mechanism to induce
any mass term which is forbidden by SUSY zero mechanism. Namely, we can
give masses to any fields which have no mass term from the superpotential
with SUSY vacua. 

In summary, we have examined a solution for the $\mu$ problem.
The point is that if SUSY is broken, the Higgs mass term, which is 
forbidden by holomorphy (the SUSY zero mechanism), must be induced.
It is interesting that the proportional coefficient is determined by
anomalous $U(1)_A$ charges.  The result is independent
on the detail of the mediation mechanism of the SUSY breaking.
 We only assume that all the SUSY breaking parameters are given as
 in eq.~(\ref{SB}).
This is a remarkable feature of this 
mechanism.

Note added: After almost finishing this work, we noticed a recent paper by 
Kitano and Okada
\cite{kitano}, in which a solution for the $\mu $ problem is discussed.
Their solution has a similar structure of our solution, but is different.
Their essential point is that
when SUSY is broken, the $R$-symmetry, which forbids the SUSY Higgs mass 
term in tree level, is also broken
\footnote{This is the same argument as in 
Ref.\cite{hempfling}, though they succeeded to omit the additional symmetry
which is introduced in Ref. \cite{hempfling}.}.
 Then the Higgs mass term can be induced
by super gravity effect. Their solution requires the R-symmetry, 
which gives
a severe constraint to the possible interactions. Actually
it cannot apply to our scenario in which R-symmetry does not exist.
On the other hand, since the mechanism discussed in this paper
 requires only the holomorphic zero (SUSY zero) mechanism, 
it can apply to our scenario\cite{maekawa}.

We would like to thank H. Nakano for pointing us to Ref.\cite{kawamura}.

\end{document}